\documentclass[aps,preprint]{revtex4}

\usepackage{graphicx}

\begin{document}

\newcommand{\wt} {\widetilde}

\title{
Scaling of Coulomb pseudo-potential in
\\ $s$-wave narrow-band superconductors
}

\author{Tae-Ho Park and Han-Yong Choi }

\affiliation{Department of Physic, Institute for Basic Science
Research,
and BK21 Physics Research Division, \\
Sung Kyun Kwan University, Suwon 440-746, Korea.}

\date{\today}

\begin{abstract}

The Coulomb pseudo-potential $\mu^*$ is extracted by fitting
the numerically calculated transition temperature $T_c$
of the Eliashberg-Nambu equation which is extended to incorporate the
narrow-band effects, that is, the vertex correction and the
frequency dependence of the screened Coulomb interaction.
It is shown that even for
narrow-band superconductors, where the fermi energy $ \epsilon_F$
is comparable with the phonon frequency $ \omega_{ph}$, the
Coulomb pseudo-potential is a pertinent parameter, and is still
given by $\mu^* = \mu/[1+\mu \ln(\epsilon_F/\omega_{ph})] $,
provided $\omega_{ph}$ is appropriately scaled.

\end{abstract}

\pacs{PACS numbers: 74.70.Wz, 74.20.Fg, 74.62.Yb}

\maketitle

%\begin{multicols}{2}

%\newpage

It is generally agreed that the superconducting properties of the
fulleride superconductors can be understood in terms of
phonon-mediated $s$-wave pairing. The relevant physical parameters
are: the logarithmically averaged phonon frequency $\omega_{ln}
\approx 0.1$ eV, the Fermi energy $\epsilon_F \approx 0.2 - 0.3$
eV, the dimensionless electron-phonon coupling constant $\lambda
\approx 0.7-0.8$, and the Coulomb repulsion $\mu=V_C (\omega=0)
N_F \approx 0.3-0.4$, where $V_C (\omega)$ and $N_F$ are,
respectively, the screened Coulomb repulsion and electronic
density of states at the Fermi level \cite{hebard,gunnarsson}.
Therefore, in fullerides we have $\epsilon_F \sim \omega_{ln}$
unlike the conventional metals where $\epsilon_F \gg
\omega_{ln}$. In this case, the Coulomb pseudo-potential $\mu^*$,
which is given by $\mu^* = \mu/[1+\mu
\ln(\epsilon_F/\omega_{ph})]$, is not much reduced from $\mu$ as
$\epsilon_F$ becomes comparable with the average phonon frequency
$\omega_{ph}$ (see below for further discussion on $\omega_{ph}$
and $\omega_{ln}$). The observations that the value of
$\ln(\epsilon_F/\omega_{ph})$ is large, typically $5-10$ in
conventional wide-band superconductors, and that the frequency
dependence of the screened Coulomb interaction is negligible in
the phonon frequency region are crucial in introducing the
concept of $\mu^*$ \cite{allen}. Both of these observations do
$not$ hold in the fullerenes because $\epsilon_F \sim
\omega_{ln}$, and, consequently, it is not clear whether the
concept of Coulomb pseudo-potential is still valid in this case.
In the present Letter, therefore, we study the Eliashberg
equation extended to narrow-band superconductors to check if the
Coulomb pseudo-potential remains a pertinent parameter as in the
wide-band superconductors, that is, if $T_c$ is determined solely
by $\omega_{ln}$, $\lambda$ and $\mu^*$, where $\mu^*$ is
independent of $\lambda$ and contains all the effects of the
Coulomb repulsion. We first obtain the superconducting transition
temperature $T_c$ by numerically solving the Eliashberg-Nambu
equation. Then, the calculated $T_c$ values were fitted with a
modified $T_c$ form which is a function of $\lambda$ and $\mu$,
in order to extract $\mu^*$. We find that even for narrow-band
superconductors the Coulomb pseudo-potential is a physically
relevant concept, and is still given by $\mu^* = \mu/[1+\mu
\ln(\epsilon_F/\omega_{ph})]$, provided that $\omega_{ph}$ is
appropriately scaled.

For wide-band superconductors, the frequency dependence of the
screened Coulomb interaction is not important. The reason is that
$V_C (\omega)$ does not change much on the phonon frequency scale
of $\omega_{ph}$ when $\epsilon_F \gg \omega_{ph}$, since the
frequency scale over which $V_C (\omega)$ varies is $\epsilon_F$
and the pairing kernel becomes negligible beyond the region of
width $\omega_{ph}$ around the Fermi surface. This does not hold
for narrow-band superconductors, and it may be important to take
$\omega$ dependence of $V_C (\omega)$ into account. We do this by
modeling the Coulomb repulsion in terms of the onsite Hubbard
repulsion and calculating its screening self-consistently.
Another complication for narrow-band superconductors is that the
Migdal theorem does not hold and the vertex correction may become
important. The vertex correction is included in the present work
following the Nambu method \cite{nambu}. These extensions, that
is, (a) frequency dependence of the screened Coulomb interaction
and (b) vertex correction, are included in the Eliashberg
equation. Gunnarsson and his coworkers pointed out that the
procedures of screening and renormalizing away the high energy
states beyond a cutoff frequency do not commute in narrow-band
superconductors \cite{gunna2}, and there arises a subtle problem
in the concept of Coulomb pseudo-potential. This problem does not
occur in the present approach. Including the full frequency
dependence of the self-consistently screened $V_C (\omega)$ in
the Eliashberg equation unnecessitates the introduction of the
cutoff frequency in manipulating $V_C (\omega)$ in the Eliashberg
$T_c$ equation \cite{schrieffer}, which becomes ill-defined in
the narrow-band superconductors.

It is straightforward to derive the Eliashberg-Nambu (EN) equation
for finite bandwidth superconductors in the Matsubara frequency
\cite{choi}:
\begin{eqnarray}
W_n  &=&  p_n +  \frac{1}{\beta}\sum_m
\frac{2\theta_m}{\sqrt{W^2_m +\phi^2_m}}
\{\lambda_+ (v_1W_m-v_2\phi_m)- \lambda_-  v_3\phi_m \},
\nonumber \\
\phi_n &=&  \frac{1}{\beta}\sum_m \frac{2\theta_m}{\sqrt{W^2_m +
\phi^2_m}}  \{\lambda_- (v_1\phi_m+v_2 W_m) + \lambda_+  v_3 W_m\},
\label{en}
\end{eqnarray}
where $p_n = 2 \pi T (n +1/2)$ is the Matsubara frequency,
$\theta_n = \tan^{-1} \left[ \epsilon_F / \left( Z_n \sqrt{p_n^2
+\Delta_n^2} \right) \right]$, and $T$ is the temperature.
$W_n \equiv p_n Z_n$ and $\phi_n \equiv \Delta_n Z_n$, where
$\Delta_n \equiv \Delta(ip_n)$ and $ Z_n \equiv Z(ip_n)$ are,
respectively, the superconducting pairing and the renormalization
functions when analytically continued into the real frequency ($
ip_n \rightarrow \omega +i \delta$), and
\begin{eqnarray}
\lambda_+ = \lambda_{ph}(n-m) - \lambda_{ch} (n-m) + \lambda_{sp}(n-m), \nonumber\\
\lambda_- = \lambda_{ph}(n-m) - \lambda_{ch} (n-m) -\lambda_{sp}(n-m).
\label{lambda}
\end{eqnarray}
Here, $\lambda_{ph} (n-m)$ is the pairing kernel due to the
electron-phonon interaction at the Matsubara frequency $(p_n -p_m
) = 2\pi (n-m) T$ given by
\begin{eqnarray}
\lambda_{ph} (n-m) &=& \int_0^{\infty} d\Omega \frac{N_F 2 \Omega
\alpha^2 F(\Omega) } {\Omega^2 +(p_n -p_m )^2 },
\end{eqnarray}
where $\alpha^2 F(\Omega)$ is
the phonon spectral function.
$ \lambda_{ch}$ and $\lambda_{sp}$ are, respectively,
the interactions in the charge and spin channels
due to the Hubbard repulsion $U$, and are determined
self-consistently as
\begin{eqnarray}
\lambda_{ch} (n-m) &=& U N_F \{ 1/2 - (\chi_n
+\chi_s) + (\chi_n +\chi_s)^2 \ln [1 +1/(\chi_n +\chi_s) ] \},
\nonumber \\
\lambda_{sp} (n-m) &=& U N_F \{ 1/2 +(\chi_n
-\chi_s)  + (\chi_n -\chi_s)^2 \ln[1 -1/(\chi_n -\chi_s)] \},
\label{kernel}
\end{eqnarray}
where $\chi_n (n-m)$ and $\chi_s (n-m)$ are the dimensionless
susceptibilities from, respectively, the normal and pairing
processes given by \cite{choi}
\begin{eqnarray}
\chi_n (k) &=&  \frac{N_F U}{ \epsilon_F} \frac{1}{\beta}\sum_l
\frac{\theta_l \theta_{k+l} p_l p_{k+l}}{\sqrt{p_l^2 +\Delta_l^2}
\sqrt{p_{k+l}^2
+\Delta_{k+l}^2} }, \nonumber \\
\chi_s (k)  &=&  \frac{N_F U}{ \epsilon_F} \frac{1}{\beta}\sum_l
\frac{\theta_l \theta_{k+l} \Delta_l \Delta_{k+l}} {\sqrt{p_l^2
+\Delta_l^2} \sqrt{p_{k+l}^2 +\Delta_{k+l}^2} }. \label{chi}
\end{eqnarray}
The finite bandwidth is explicitly incorporated through the factor $\theta$,
which is equal to $\pi/2$ in the conventional infinite bandwidth case.
Finally, the vertex corrections are included by extending the
Nambu method \cite{nambu}. The vertex functions are given by
\begin{eqnarray}
v_1 &=& \frac{W_n - W_m }{p_n -p_m},
\nonumber \\
v_2 &=& \phi_n \frac{p_n -p_m }{(p_n -p_m)^2+\omega^2_{pl}},
\nonumber \\
v_3 &=& \phi_m \frac{p_n -p_m } {(p_n -p_m)^2+\omega^2_{pl}},
\label{vertex}
\end{eqnarray}
where $\omega_{pl}$ is the plasma frequency which is taken to be
1 eV. Without the vertex correction, which corresponds to $v_1=1$
and $v_2 = v_3 =0$, Eq.\ (\ref{en}) is of the same form as in the
theory used to study spin fluctuation effects on
superconductivity \cite{berk}. Self-consistent solution of Eq.\
(\ref{en}) together with Eqs.\ (\ref{lambda})--(\ref{vertex})
yields $Z(ip_n)$ and $\Delta(ip_n)$ in the Matsubara frequency.
Details of the formulation will be reported elsewhere
\cite{choi2}.

It is simple to calculate $T_c $ from the EN equation of Eq.\
(\ref{en}). It is identified as the highest temperature at which
the pairing function $\Delta$ is non-vanishing. The input parameters
for solving the coupled EN equations are: $\alpha^2 F(\Omega)$
(which determines $\omega_{ln}$ and $\lambda$), $U$, and
$\epsilon_F$.
Note that the long range part of the Coulomb interaction $V_q$ is not included
in the present study.
This may be justified for the case where the average of $V_q$ is small
compared with the onsite Hubbard repulsion $U$:
$ \left< V_q \right> _{ave}/U \approx 4 \pi n e^2/k_F^2 U  \ll 1$.
Alternatively, $U$ may be considered as a partially screened value due
to the metallic screening of the long range part of the Coulomb interaction.
The $\omega_{ln}$ is given by
\begin{eqnarray}
\omega_{ln} = \exp \left[ \frac{2}{\lambda} \int_0^{\infty}
d\Omega \ln \Omega \frac{1}{\Omega} N_F \alpha^2 F (\Omega)
\right],
\end{eqnarray}
where $\lambda = N_F V$, $\mu=U N_F$, and $V=2 \int_0^{\infty}
d\Omega \alpha^2 F(\Omega)/\Omega$. We take $\alpha^2 F(\Omega)$
as a sum of three truncated Lorentzians appropriate for fullerene
superconductors \cite{choi}. This choice of $\alpha^2 F(\Omega)$
gives the logarithmically averaged phonon frequency $\omega_{ln}
= 0.094$ eV. $\mu$ is varied between 0 and 0.4 and $\lambda$
between 0.4 and 1.2 to calculate $T_c$, from which to extract an
extended $T_c$ formula and the Coulomb pseudo-potential $\mu^*$.

The EN equation was solved by iterations, for various values of
$\lambda$ and $\mu$ to find $T_c$. Self-consistency is achieved
with a few tens of iterations except when $T$ is close to $T_c$.
These results are used to extract a modified $T_c$ formula
following the same procedure as McMillan and Allen (MA) \cite{ma}.
The MA $T_c$ equation is given by
\begin{eqnarray}
T_c = \frac{\omega_{ln}}{1.2} \exp \left[ -
\frac{1.04(1+\lambda)}{\lambda-\mu^* (1+0.62 \lambda)} \right],
\label{ma}
\end{eqnarray}
which was extracted by fitting to numerical $T_c$ calculations of
the Eliashberg equation. The McMillan-like procedure for
extracting $T_c$ formula proceeds as follows \cite{ma}: Firstly,
for the case of zero Coulomb potential, $\mu = \mu^* = 0$, we
estimate the $\lambda$ dependence of $T_c$ by taking
\begin{equation}
\ln \left[ \frac{\omega_{ln}}{T_c} \right] = a +\frac{b}{\lambda}
+\frac{c}{\lambda^2},
\label{log}
\end{equation}
where $a$, $b$, and $c$ are the constants to be determined by
fitting to the exact numerical results obtained by solving EN
equation of Eq.\ (\ref{en}). The MA case corresponds to $a,b \neq
0$ and $c=0$. Non-zero $c$ introduces higher power of $1/\lambda$
in the $T_c$ formula due to the narrow bandwidth effects
considered here, that is, the vertex correction and frequency
dependence of the effective Coulomb interaction. McMillan used an
approximate solution of the Eliashberg equation to guide the
choice of the functional form of Eq.\ (\ref{log}). From a solution
of a version of a square-well potential model, the form of $ \ln
( \omega_{ln}/ T_c ) = a + b/\lambda $ was anticipated \cite{ma}.
The simplest correction coming from beyond the simple model will
be of the form $1/\lambda^2$ as incorporated in Eq.\ (\ref{log}).
For the parameter values appropriate for the fulleride
superconductors, we obtain $a = 1.13$, $b = 0.765$, and $c =
0.033$. This should be compared with the MA case, $a=1.22$,
$b=1.04$, and $c=0$. The obtained values for $a,~b$, and $c$
depend weakly on the particular choice of
$\epsilon_F/\omega_{ln}$ and $\alpha^2F$.
%As we increase
%$\epsilon_F/\omega_{ln}$, the calculated values of $a,~b$, and
%$c$ approach the MA values.

\begin{figure}[t!]
\begin{center}
\includegraphics*[scale=0.3]{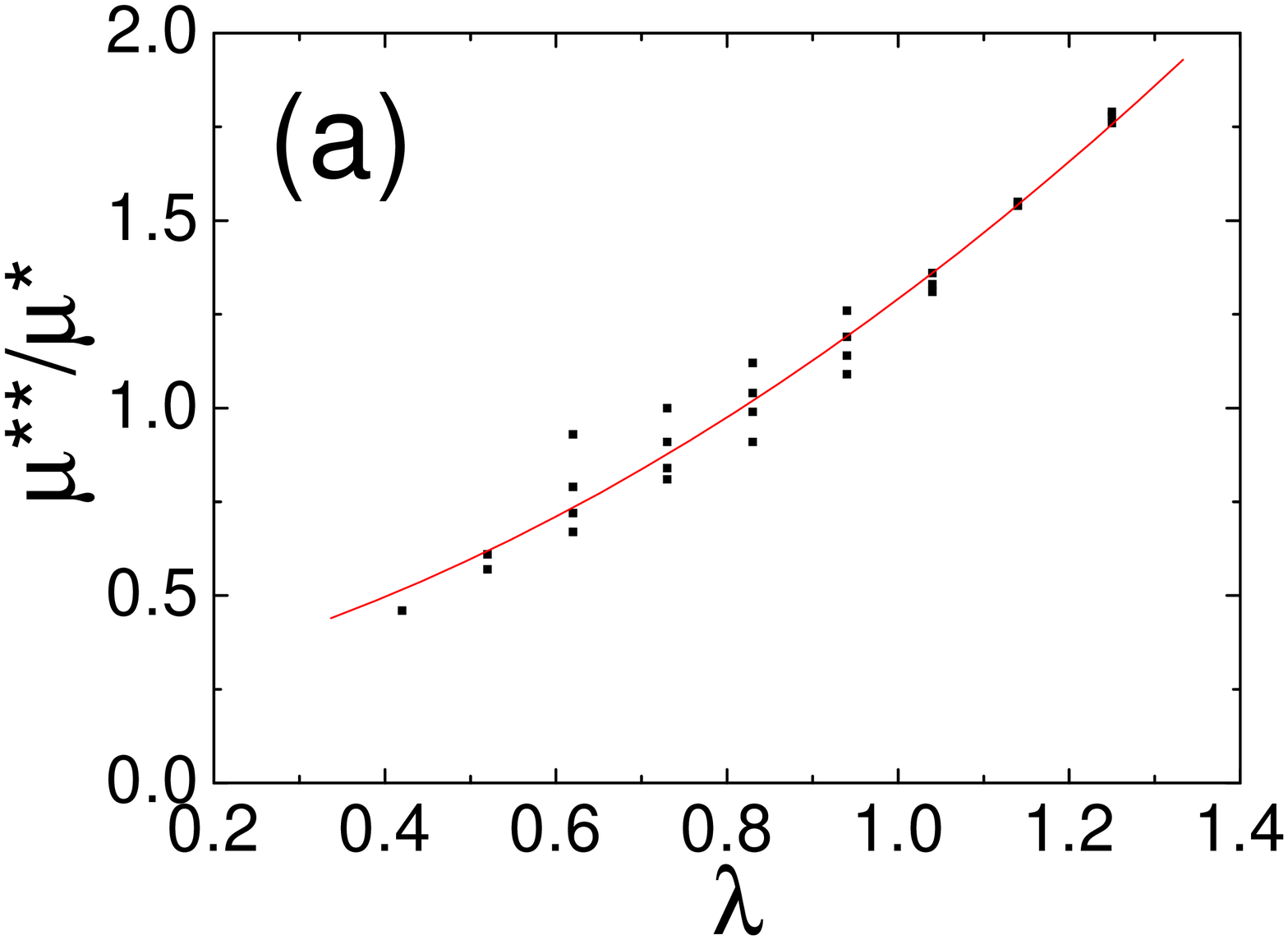}
%\epsf{figure=fig-1a.eps,width=0.9\linewidth}
\includegraphics*[scale=0.3]{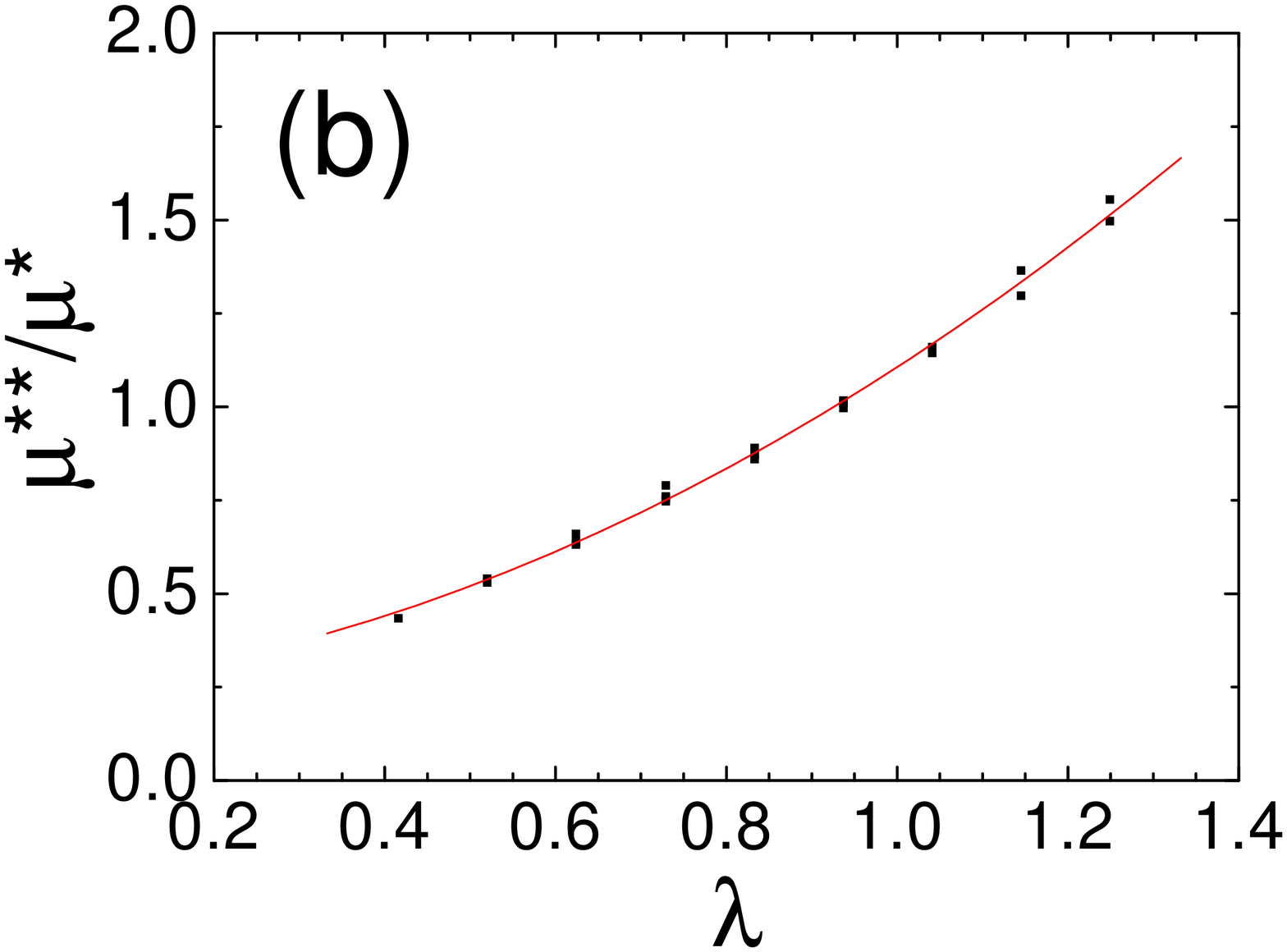}
%\epsf{figure=fig-1b.eps,width=0.9\linewidth}
\vspace{0.2in}
\end{center}
\caption{Determining $S=\omega_{ph}/\omega_{ln}$ by requiring
$\mu^{**}/\mu^*$ be independent of $\mu$, when the bandwidth $B$
is 0.5 eV. $\lambda$ is varied from 0.4 to 1.2, and, for each
$\lambda$, four $\mu$ values, 0.1, 0.2, 0.3, and 0.4, were taken
to calculate $\mu^{**}/\mu^*$. (a) When $S=1$. It is clear that
for each $\lambda$, the four calculated points corresponding to
four different $\mu$ values are scattered in the plot. (b) When
$S=2$, the scattered 4 points in the figure (a) scale to collapse
on to a single point for each $\lambda$. }
\end{figure}

Secondly, to include the effects of non-zero Coulomb interaction
on $T_c$ ($\mu \neq 0$), we take the form
\begin{eqnarray}
\ln \left[ \frac{ \omega_{ln}} {T_c} \right] =
\frac{a\lambda^2+b\lambda+c}{\lambda^2-\mu^{**} (\mu,\lambda)}.
\label{tc}
\end{eqnarray}
as with MA. Here, $a$, $b$, and $c$ have already been determined
from the first fitting process, and $\mu^{**}$ depends on $\mu$
and $\lambda$. For the MA case, $\mu^{**}=\mu^*(1+0.62 \lambda)$,
where
\begin{equation}
\mu^* = \mu/[1+\mu \ln(\epsilon_F /\omega_{ph} )] \label{mustar}
\end{equation}
as given by Morel and Anderson \cite{morel}. We put
\begin{equation}
\mu^{**} \equiv \mu^* f(\lambda), \label{mu2star}
\end{equation}
where $f(\lambda)$ is a function of $\lambda$ to be determined by
fitting, and $\mu^*$ is assumed, in the present case as well, to
take the same Coulomb pseudo-potential form of Eq.\ (\ref{mustar}).
Then, $f(\lambda)$ can be written as
\begin{equation}
f(\lambda) = \frac{1+ \mu \ln( \epsilon_F/\omega_{ph} )} {\mu}
\left[ \lambda^2- \frac{(a\lambda^2+b\lambda+c) } {
\ln(\omega_{ln}/T_c)} \right].
\end{equation}
We then write
\begin{equation}
\omega_{ph} = S \omega_{ln},
\end{equation}
where $S$ is to be determined by requiring that $\mu^{**}/\mu^*
=f(\lambda)$ be $independent ~of~ \mu$. We calculate, for a fixed
$\lambda$, $\mu^{**}/\mu^*$ for several values of $\mu$.
$\mu^{**}/\mu^*$, in general, will depend on $\mu$ and be
scattered in the plot as shown in Fig.\ 1(a) for $S = 1$. For an
appropriate value of $S$, however, the data will scale to
collapse on to a single point for each $\lambda$. This is found to
occur at $S =2$ when $\epsilon_F=0.25$ eV as can be seen in Fig.\ 1(b).
Fitting the curve of Fig.\ 1(b) yields $f(\lambda) =
\mu^{**}/\mu^* = 0.618 \lambda^2 +0.244 \lambda +0.244 $.

\begin{figure}[t!]
\includegraphics*[scale=0.3]{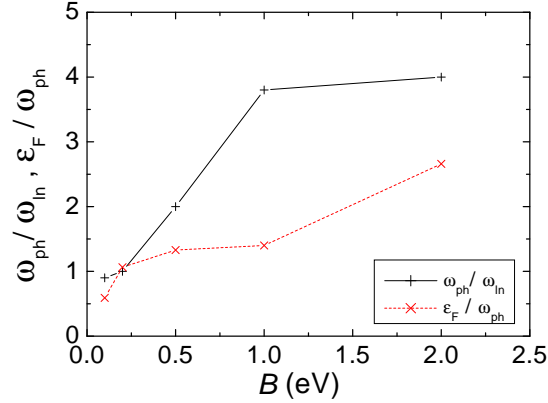}
%\epsf{figure=fig-2.eps,width=0.9\linewidth}
\vspace{0.2in} \caption{$S =\omega_{ph}/\omega_{ln}$ as a function
of the bandwidth $B$. $S$ was determined by requiring
$\mu^{**}/\mu^*$ be independent of $\mu$ for a given bandwidth.
The + and $\times$ are, respectively, calculated values for $S$
and $\epsilon_F /\omega_{ph}$. The solid and dashed lines are a
guide to eyes.}
\end{figure}

This scaling behavior gives reliability to our procedures.
Consequently, $T_c$ can be
written as
\begin{equation}
T_c = \omega_{ln} \exp \left[ - \frac{a \lambda^2 +b \lambda +c}
{\lambda^2 -(a' \lambda^2 +b' \lambda +c' ) \mu^*}. \right]
\label{newtc}
\end{equation}
MA case corresponds to $c=c'=0$. The concept of Coulomb
pseudo-potential is found to be valid and important even for a
narrow-band superconductor of $\epsilon_F/\omega_{ph} \sim 1$
where retardation effects are much weaker compared with the
conventional wide-band superconductors. We should note here that
although we determined $S=\omega_{ph}/\omega_{ln}$
by requiring $\mu^{**}/\mu^*$ be independent of $\mu$,
$\omega_{ph}$ should $not$ be considered an additional
fitting parameter. $S$ remains the same for a given bandwidth
irrespective of $\lambda$ and $\mu$, and gives the appropriate
Coulomb pseudo-potential value $\mu^*$ for given $\mu$ and
bandwidth.

We then vary the bandwidth $B = 2\epsilon_F $ to find how $S$
varies accordingly. For each $B$, we repeated the fitting
procedures described above to determine
$S=\omega_{ph}/\omega_{ln}$ which gives the best scaling
behavior. $S$ as a function of $B$ determined this way is shown
in Fig.\ 2. $S$ increases as the bandwidth is increased, and then
saturates around the value of 4 when $B$ is larger than 1 eV for
$\omega_{ln} = 0.094$ eV. Note that in the Eliashberg theory, the
Coulomb pseudo-potential $\mu^*$ of Eq.\ (\ref{mustar}) is often
rescaled with a cutoff frequency $\omega_c \approx 5-10~
\omega_D$, where $\omega_D$ is the Debye frequency. $\omega_c$
corresponds to $\omega_{ph}$ in the present work. The value of $S
\approx 4$ for $B>1$ is consistent with the $\omega_c /\omega_D
\approx 5 $ often adapted for conventional wide-band
superconductors \cite{allen}. Increase of $S$ as $B$ is increased
does not mean the less effective retardation, because the
increase of $S$ is slower than that of $\epsilon_F$. This is
shown with the crosses in the plot. $\epsilon_F /\omega_{ph}$
increases as $B$ is increased, and the retardation, which is
determined by $\epsilon_F /\omega_{ph}$ as given by Eq.\
(\ref{mustar}), is more effective for wide-band superconductors.

To summarize, we investigated whether the Coulomb pseudo-potential
$\mu^*$ is still a pertinent concept in the narrow-band
superconductors. Because the narrow bandwidth implies a
negligible retardation effects ($\mu^* \approx \mu $), it is not
clear whether the $\mu^*$ remains a relevant and useful parameter
for the narrow-band superconductors like the fullerenes. To answer
this question, we solved the extended Eliashberg equation which
incorporates the finite bandwidth effects, namely, the frequency
dependence of the effective interaction between electrons and the
vertex corrections, and obtained the transition temperature $T_c$
as the dimensionless electron-phonon coupling constant $\lambda$
and the Coulomb repulsion $\mu$ are varied. The obtained
numerical results were used to find a $T_c$ formula which is
valid for narrow-band superconductors and to extract the Coulomb
pseudo-potential $\mu^*$. It was found that $T_c$ is given by the
form of Eq.\ (\ref{newtc}) where $\mu^*$ contains all the effects
of the Coulomb repulsion and is independent of $\lambda$.
Therefore, the concept of Coulomb pseudo-potential continues to be valid
even for narrow-band superconductors. Another
interesting and important outcome of the present study is that the
Coulomb pseudo-potential is still given by $\mu^* = \mu/[1+\mu
\ln(\epsilon_F/\omega_{ph})]$ as in the conventional wide-band
superconductors, provided $\omega_{ph} =S \omega_{ln}$ is
appropriately scaled. Scaling of $\omega_{ph}$ and $\mu^*$ as
$\epsilon_F$ is varied is also computed. Because we included the
vertex correction and frequency dependent Coulomb interaction in
the present work, there is no {\it a priori} reason that the
Coulomb pseudo-potential take the same form. We established that
the Coulomb pseudo-potential continues to be a pertinent
parameter and is still given by the Morel-Anderson form in the
narrow-band superconductors.

We would like to thank Yunkyu Bang for helpful conversations.
This work was supported by Korea Science \& Engineering
Foundation (KOSEF) through Grant No.\ 1999-2-114-005-5, by Korea
Research Foundation (KRF) through Grant No.\ KRF-2000-DP0139, and
by the Ministry of Education through the Brain Korea 21 SNU-SKKU
Physics Program.

%\end{multicols}

\end{document}